# A Review of ChatGPT Applications in Education, Marketing, Software Engineering, and Healthcare: Benefits, Drawbacks, and Research Directions

Mohammad Fraiwan [1,*] and Natheer Khasawneh [2]

[1] Department of Computer Engineering, Jordan University of Science and Technology, P.O. Box 3030, Irbid 22110, Jordan; mafraiwan@just.edu.jo

[2] Department of Software Engineering, Jordan University of Science and Technology, P.O. Box 3030, Irbid 22110, Jordan; natheer@just.edu.jo

*Correspondence: mafraiwan@just.edu.jo

**Abstract:** ChatGPT is a type of artificial intelligence language model that uses deep learning algorithms to generate human-like responses to text-based prompts. The introduction of the latest ChatGPT version in November of 2022 has caused shockwaves in the industrial and academic communities for its powerful capabilities, plethora of possible applications, and the great possibility for abuse. At the time of writing this work, several other language models (e.g., Google's Bard and Meta's LLaMA) just came out in an attempt to get a foothold in the vast possible market. These models have the ability to revolutionize the way we interact with computers and have potential applications in many fields, including education, software engineering, healthcare, and marketing. In this paper, we will discuss the possible applications, drawbacks, and research directions using advanced language Chatbots (e.g., ChatGPT) in each of these fields. We first start with a brief introduction and the development timeline of artificial intelligence-based language models, then we go through possible applications of such models, after that we discuss the limitations and drawbacks of the current technological state of the art, and finally we point out future possible research directions.

## 1 Introduction

ChatGPT is a type of artificial intelligence (AI) language model that uses deep learning algorithms to generate human-like responses to text-based prompts. The introduction of the latest ChatGPT version in November of 2022 has caused shockwaves in the industrial and academic communities for its powerful capabilities, plethora of possible applications, and the great possibility for abuse. At the time of writing this work, several other language models (e.g., Google's Bard and Meta's LLaMA) just came out in an attempt to get a foothold in the vast possible market. These models have the ability to revolutionize the way we interact with computers and have potential applications in many fields, including education, software engineering, healthcare, and marketing.

Historically, language models have been around for more than 20 years with some attempts go back to the 1960's. However, recent developments in deep learning AI, the huge computational power offered by graphical processing units (GPUs), and the accessibility to large datasets have enabled amazing advancements in the capabilities and the likelihood to human operators. Moreover, Chatbots are now being touted as the future of search engines, because they are able to formulate answers to queries rather than just point out the links to possible answers. For example, instead of searching for programming tutorials or a lucky similar answer to a homework assignment, ChatGPT can easily provide the necessary code in response to such a query with varying degrees of complexity. ChatGPT was able to pass bar exams, United States medical licensing exams, and job interviews, among others [1].

The first language models appeared in the 1960s. ELIZA was developed by Joseph Weizenbaum as one of the first Chatbot programs [2]. It used pattern matching and pre-written responses to simulate conversation with a human user. Fast forward to the 1990s, the artificial linguistic Internet computer entity (ALICE) was developed by Richard Wallace. ALICE was another early Chatbot program. It used a similar approach to ELIZA, but also incorporated machine learning to improve its responses over time [3]. A decade later, Cleverbot was introduced by Rollo Carpenter [4]. Cleverbot was a Chatbot program that used artificial neural networks to learn from its conversations with users. It was able to generate more natural and varied responses than earlier Chatbots. More recently, in 2018, OpenAI released the first version of their Generative Pre-trained Transformer (GPT) language model. It used unsupervised learning to train on large amounts of text data and could generate coherent and diverse text based on a given prompt. In 2019, OpenAI released an improved version of their GPT model called GPT-2. It had 1.5 billion parameters, making it one of the largest language models at the time. GPT-2 was able to generate high-quality text that was difficult to distinguish from human writing. In 2020, OpenAI released an even more powerful version of their GPT model called GPT-3. It had 175 billion parameters, making it the largest language model to date. GPT-3 was able to perform a wide range of language tasks, including language translation, content generation, and answering questions [5]. Moreover, the number of users of its service was reported by media outlets to exceed 100 Million in two months after its launch. The latest release, ChatGPT-4 is scheduled for release in mid March, 2023. Moreover, at the time of writing this paper, Microsoft just released Visual ChatGPT, which extends the capabilities of ChatGPT by allowing sending/receiving images in the chat dialogue.

ChatGPTs are built on top of the GPT models developed by OpenAI, but with additional training and customization for conversational applications. They represent the cutting edge of AI language technology and have the potential to revolutionize the way we interact with computers and each other. On February 6th, 2023, Google releases their own conversational AI called Bard, which was quickly followed by Meta's Large Language Model Meta AI (LLaMA) on February 24th, 2023. Other less known models do exist in the literature, including bidirectional, extreme multilingual language model (XLNet) [6], GShard (a Transformer-based deep learning architecture) [7], robustly optimized BERT pretraining approach (RoBERTa) [8], and text-to-text transfer transformer (T5) [9]. These AI language models are all based on the transformer architecture and have achieved impressive results in various natural language processing tasks. However, each model has its own strengths and weaknesses, and the most appropriate model depends on the specific application and the data available.

These powerful language models represent a technological disruption to the current academic and industrial landscape. They may render many existing technologies (e.g., traditional search engines) obsolete/insufficient. Moreover, they may adversely affect the educational paradigm with the way current assignments and evaluations are performed. On the other hand, it may open further avenues for exploration and learning if used properly. In this paper, we discuss the effects that the introduction of sophisticated have on education, software engineering, healthcare, and marketing. The applications, drawbacks, and possible research directions are presented in the next few sections. However, the effects of the latest Chatbot language models are still being felt and more applications/drawbacks are coming up every day.

The remainder of this paper is organized as follows. In section 2 we present the possible applications of ChatGPT in the four identified fields. The drawbacks are discussed in section 3. Future research directions are explored in section 4. We conclude in section 5.

## 2  Applications of ChatGPT and Language Models

In this section, we highlight the possible applications of ChatGPT, as well as other advanced language models being rolled out, in the four aforementioned fields. As more people use and adopt these AI tools, other avenues are possible and this is an ongoing and evolving topic.

### 2.1  Education

Language models has several applications in education, such as providing personalized learning experiences, generating test questions and answers, and facilitating online discussions. It can also assist teachers in grading assignments and providing feedback to students. Some of the activities involved include:

• Language models can assist in providing personalized learning experiences by analyzing student performance data and generating adaptive learning pathways. It can recommend appropriate learning materials, answer students' questions, and provide feedback on assignments.

• Generate test questions and answers for students, which can save time for teachers and ensure that tests cover a range of topics and levels of difficulty.

• Facilitate online discussions between students and teachers by generating conversation prompts, answering questions, and providing feedback on responses. This can enhance collaboration and engagement in online learning environments.

• Assist teachers in grading assignments and providing feedback to students. It can identify areas where students need improvement and suggest ways to improve their work.

• Assist language learners by generating exercises, providing pronunciation feedback, and answering questions about grammar and vocabulary. It can also provide conversational practice for language learners by generating dialogue prompts.

• Assist special education students by generating alternative learning materials, providing additional explanations, and answering questions in a way that is tailored to their individual needs.

Overall, ChatGPT's education applications have the potential to enhance student learning experiences, provide teachers with valuable resources and assistance, and improve the efficiency and effectiveness of online learning environments. Tapalova and Zhiyenbayeva [10] recently explored the possibilities of AI in education. A survey of educators at their institute indicated that education can be made more effective with the help of AI. More specifically, AI can facilitate personalization of the educational activities, increase availability of resources, improve adaptability of the educational material to individual student needs, provide prompt and continuous feedback, and improve mental motivation and stimulation. However, it is essential to ensure that ChatGPT is used responsibly and thoughtfully, with considerations for potential biases and ethical concerns. In another study, Kumar and Boulanger [11] explored the use of deep learning AI to automatically grade essays using rubric instead of holistic scores. They concluded that it is possible for language models to aid students in learning proper writing and its strategies.

Language models can be a useful tool for teaching and learning, particularly in the field of language arts and writing. Here are a few ways that language models can be used to enhance teaching:

• Writing prompts: Language models can be used to spark students' creativity and engage them in writing. For example, enter a topic or theme and ask the Chabot to generate a writing prompt for students to work on.

• Writing feedback: After students have written a piece, the language model can provide feedback on their work. For example, ChatGPT can analyze the writing for grammar, punctuation, and spelling errors, as well as provide suggestions for improving the overall structure and flow of the writing.

• Language practice: Advanced language models can help students practice their language skills. For example, one can ask ChatGPT to provide synonyms or antonyms for certain words, or to provide sample sentences using certain grammar structures or vocabulary.

• Research assistance: language models can be used to assist students in their research by acting as an advanced search engine. However, this disrupts the current homework/assignment models with the lack of ability to detect plagiarism using ChatGPT.

• Personalized learning: Language models can also be used to create personalized learning experiences for students. For example, ChatGPT can be used to provide feedback and guidance to individual students based on their specific strengths and weaknesses in writing.

However, assessing the impact of using ChatGPT in teaching and learning is important to determine whether it is an effective tool for improving students' skills. The following are a few ways to measure the impact of using ChatGPT in teaching:

• Pre- and post-assessments to measure the improvement in students' skills. The assessments should be aligned with the learning objectives and outcomes of using ChatGPT. For example, assess students' writing skills before and after using ChatGPT to see if there is an improvement in the quality of their writing.

• Analyzing student work to see if there is an improvement in their skills. A rubric to assess students' writing, and compare their work before and after using ChatGPT to see if there is an improvement in areas such as grammar, sentence structure, and vocabulary.

• Student feedback. Surveys or focus groups can be used to gather their feedback on the usefulness of ChatGPT in improving their skills, as well as their overall experience of using the tool.

• Observations to assess student's engagement and level of participation. This involve observing their interactions with ChatGPT, their level of motivation, and their ability to use the tool effectively.

• Comparison with control group. If possible, the progress of students who used ChatGPT can be compared with a control group of students who did not use the tool. This can help to isolate the impact of using ChatGPT and determine whether it was a significant factor in improving students' skills.

By using these methods to assess the impact of using language models in teaching, it is possible to determine whether it is an effective tool for improving students' skills, and make any necessary adjustments to the teaching methods to further enhance learning outcomes.

### 2.2 Software Engineering

ChatGPT can be used in software engineering for tasks such as generating code, debugging, and software testing. It can also help developers in natural language processing tasks, such as analyzing user requirements, and generating user interfaces. This can be accomplished as follows:

• Code generation: Generate code snippets for developers based on natural language descriptions of the desired functionality. This can save time and improve efficiency in the software development process.

• Debugging: Assist in debugging code by identifying errors and suggesting fixes based on natural language descriptions of the issue.

• Software testing: generate test cases and test data based on natural language descriptions of the desired test scenarios. This can improve the efficiency and effectiveness of software testing.

• Natural language processing: Assist developers in natural language processing tasks, such as analyzing user requirements, generating user interfaces, and providing Chatbot interactions with users.

• Documentation generation: Generate software documentation based on natural language descriptions of the software's functionality. This can save time for developers and improve the quality and completeness of the documentation.

• Collaboration and knowledge sharing: Facilitate collaboration and knowledge sharing between developers by generating conversation prompts, answering questions, and providing feedback on responses. This can enhance communication and efficiency in software development teams.

Overall, ChatGPT's software engineering applications have the potential to improve efficiency and effectiveness in the software development process, enhance collaboration and knowledge sharing, and improve the quality of software documentation. Raychev et al. [12] identified early on the potential of natural language processing in synthesizing code completion and predicting the probability of sentences. In addition, they used a similar approach to predict syntactic and semantic variable types and identified names. Such efforts have resulted in a slew of studies that research the role of language models in specific software engineering problems [13]. In an another study, Tu et al. [14] investigated the role of language models in predicting the repetitive, regular, and typical code snippets in human-written programs. This has the potential to improve code suggestions in automatic completion systems. Furthermore, Allamanis et al. [15] investigated the role of language models in detecting software bugs [16]. In another avenue, language models have been shown to useful in the development and update of software comments and documentation. These efforts show that it is possible to tap into the power of language models in the software engineering domain.

## 2.3 Healthcare

ChatGPT can assist healthcare professionals by providing patient triage, symptom analysis, and medical diagnosis. It can also aid in drug discovery and clinical trials. This can be accomplished as follows:

- Patient Care: Assist healthcare professionals by providing personalized care to patients. It can answer questions about medical conditions, treatments, and medications, and provide recommendations based on patient symptoms and medical history.
- Electronic health records (EHR): Assist in updating electronic health records by analyzing natural language descriptions of patient conditions and treatments and generating corresponding entries in the EHR.
- Medical education: Assist medical students and professionals by generating medical case studies, answering questions about medical terminology, and providing educational resources for medical training.
- Mental health: Assist in mental health care by providing personalized support and resources to patients. It can answer questions about mental health conditions, provide coping strategies and relaxation techniques, and offer support for patients experiencing mental health crises.
- Clinical trials: assist in clinical trials by generating eligibility criteria, screening questions, and informed consent forms based on natural language descriptions of the trial's objectives.
- Telemedicine: Assist in telemedicine by facilitating communication between healthcare professionals and patients. It can answer questions about telemedicine procedures, provide technical support for patients using telemedicine tools, and assist in scheduling appointments.

In general, ChatGPT's healthcare applications have the potential to improve patient care, enhance medical education and training, and improve efficiency in healthcare operations. Adlung et al. [17v] identified two challenges facing artificial intelligence in clinical decision making, mainly explainability and causability. Explainability refers to the ability of the model to provide results that can be justified (e.g., significant factors that have statistical association with the output). Although examples on the Web were able to show ChatGPT giving reasons for their answers or the lack of a correct answer, language models still has the potential to open up deep learning models for easy explanations and transparency. This also relates to causability, from a legal and regulatory perspective, adopting artificial intelligence algorithms in clinical decision making may require such methods to provide clear explanations on why a certain output was generated. Moreover, Wu et al. [18] pointed out the importance of literature review in the field of regulatory science and the role that language models can play in accelerating this review. These factors bring language models to the fore of the medical AI domain. Lederman [19] discussed clinical natural language processing (cNLP) that use textual data in health records to support the clinical decision making process. They argued for a rethink of cNLP systems to improve their practical adaptation and deployment [20]. Specifically, they identify several factors that hinder the usability of the cNLP systems, which may be overcome by advanced language models, including handling of complex language processing, ability to answer "how" and "why" questions, and the problem of explainability. In an interesting study, Liu et al. [21] investigated the role of language models in the development and discovery of drugs. Mainly, language models can have a great role in the quick identification of targets, optimization of clinical trials, facilitation of decision making from a regulatory perspective, and promoting pharmacovigilance [22]. Bhatnagar et al. [23] also reviewed the role of language models in discovering new drugs, clinical trials, and pharmacovigilance.

The topic of pharmacovigilance is another interesting are, where language models can gauge, analyze, and detect drug-related problems or adverse effects/interactions from users' prompts and discussions [24]. Ball and Pan [25] explored the use of language models in the processing of the individual case safety reports submitted to the Food and Drug Administration as part of their adverse event reporting system. They identified several problems that need to be resolved in order to facilitate the acceptance of language models in pharmacovigilance. Koneti et al. [26] proposed using language models in drug development by extracting unstructured data from pharmacokinetics and pharmacodynamics study reports. Several language models were proposed recently for the purpose of medical text mining. Wang et al. [27] developed the "DeepCausality" model, which is able to include AI language models in order to create a causal inference model from fee text. They demonstrated its effectiveness in detecting idiosyncratic drug-induced liver injury with high accuracy. Lee et al. [28] proposed bidirectional encoder representations from transformers for biomedical text mining (BioBert) model, which was able to answer biomedical questions, extract relations, and recognize named biomedical entities with improved accuracy. Similarly, ClinicalBert [29,30] was proposed to predict hospital readmissions using medical text data from hospital admission notes and discharge summaries.

### 2.4 Marketing

ChatGPT can be used in marketing to generate product descriptions, customer reviews, and social media content. It can also assist in Chatbot interactions with customers and provide personalized recommendations based on customer preferences. The following are some elaborations on ChatGPT's marketing applications:

• Content creation: Assist in content creation for marketing campaigns by generating ideas for social media posts, email marketing campaigns, and blog articles. It can also generate headlines, product descriptions, and promotional messages based on natural language descriptions of the marketing objectives.

• Customer service: Assist in customer service by providing personalized support to customers. It can answer questions about products and services, provide recommendations based on customer preferences, and assist in resolving customer issues and complaints.

• Lead generation: Assist in lead generation by analyzing customer data and generating natural language descriptions of potential customers. It can also assist in lead qualification by analyzing customer responses and identifying potential leads.

• Market research: Assist in market research by generating surveys, analyzing customer feedback, and identifying trends and insights based on natural language descriptions of the research objectives.

• Personalization: Assist in personalizing marketing campaigns by analyzing customer data and generating personalized recommendations for products and services. It can also assist in tailoring marketing messages to specific customer segments based on natural language descriptions of the customer demographics and preferences.

• Sales: Assist in sales by generating personalized product recommendations and assisting in the sales process. It can also assist in upselling and cross-selling by analyzing customer data and generating natural language descriptions of potential add-ons or upgrades.

ChatGPT's marketing applications have the potential to improve efficiency and effectiveness in marketing campaigns, enhance customer experience and satisfaction, and improve sales performance. Verma et al. [31] discussed the role of recent disruptive technologies, mainly AI, in business operations. One of the areas that they have identified was the use of Chatbots and language models to improve customer experience [32] and customer relationship management (CRM) systems. Language models and Chatbots offer great advantages in the form of faster and automated access to data, simpler and efficient processes, accuracy, and cost effectiveness [33]. Similarly, De Mauro et al. [34] published a recent taxonomy of the use of machine learning and AI in marketing. The authors have identified several use cases of AI in marketing and divided those into customer side versus business side. On the customer facing side, they identified personalization of offers, communication, recommendations, and assortments as candidates for improvements. Moreover, they indicated that the consumption experience can also be improved via experience improvement and digital customer service. On the business side, machine learning can be beneficial in market understanding and customer sensing, among other avenues. In a recent literature review of marketing and AI, Duarte et al. [35] have identified recommender systems and text analysis as promising areas of useful Chatbots usage in marketing. De Bruyn et al. [36] investigated the opportunities and pitfalls of using AI in marketing. They have identified several risks associated with adopting new AI disruptive technologies, mainly bias, explainability, control, and unsafe/unrealistic learning environments. Moreover, they conclude with a warning of the possibility of AI failure in this domain if these challenges are not resolved by the implicit marketing knowledge transfer to AI models.

## 3 Drawbacks of Language Models

Although language models and Chatbots clearly offer great opportunities, they have inherent shortcomings that limit their applicability, adoption, and usefulness. In the next few subsections, we go through these drawbacks in detail.

### 3.1 Bias

Language models can exhibit bias if the training data used to create them is biased. As Schramowski et al. [37] pointed out, large pre-trained models that try to mimic natural languages, may end up repeating the same unfairness and prejudices. This can lead to discriminatory or inaccurate analyses and recommendations. Moreover, this may lead to public outcry (i.e., political, social, and legal) against the commercial applications. These biases manifests themselves in several ways, as follows:

- Training data bias: Language models are typically trained on large datasets of human language. If these datasets are biased in some way (e.g., based on race, gender, socioeconomic status, etc.), then the model may learn and replicate these biases in its responses. For example, if the training data is biased towards a particular gender, then the model may generate responses that are more favorable towards that gender.
- User interaction bias: The responses generated by Chatbots are based on the input they receive from users. If users consistently ask biased or prejudiced questions, then the model may learn and replicate these biases in its responses. For example, if users frequently ask questions that are discriminatory towards a particular group, then the model may generate responses that perpetuate these biases.
- Algorithmic bias: The algorithms used to train and operate language models and Chatbots may also introduce biases. For example, if the model is trained to optimize for a particular metric (e.g., accuracy, engagement, etc.), then it may prioritize generating responses that optimize for that metric, even if those responses are biased in some way.
- Contextual bias: Chatbots generate responses based on the context they receive from users. If the context is biased in some way (e.g., based on the user's location, language, etc.), then the model may generate biased responses. For example, if a user is asking questions about a particular culture or religion, and the model is not trained on that culture or religion, it may generate biased responses due to its lack of knowledge.

It is important to note that bias in language models are not necessarily intentional or malicious. Although this sometimes is hard to prove or justify to the non-technical public. Moreover, it can have harmful consequences, such as perpetuating stereotypes, reinforcing discriminatory attitudes, or excluding certain groups from access to information and resources. To mitigate these risks, it is of paramount importance to train and operate the models in a responsible and ethical manner, by carefully selecting and monitoring training data, incorporating diversity and inclusion considerations, and regularly auditing the model for potential biases.

### 3.2 Lack of Transparency

Language models, and deep learning models in general, are called "black box" models as their results can be difficult to interpret and understand, making it challenging for researchers to assess their validity and accuracy [38]. These models lack transparency, meaning that it is often difficult to understand how the model arrived at a particular output or decision. This lack of transparency can be problematic for several reasons:

1. Debugging: If the model generates unexpected or incorrect output, it can be challenging to identify the source of the problem without understanding how the model arrived at its decision.
2. Accountability: In some cases, the output generated by the model may have significant consequences for individuals or society as a whole (e.g., in healthcare or criminal justice). If the model lacks transparency, it can be difficult to hold it accountable for its decisions.

3. Bias: As mentioned earlier, language models can be biased in various ways, such as in the training data or algorithms used. Without transparency, it can be difficult to identify and correct these biases.

4. Trust: In many cases, users may be hesitant to trust the output generated by the model if they don't understand how it arrived at its decision. This have ramifications in obtaining regulatory approvals and adoption by the public.

### 3.3 Explainability

Researchers are developing new techniques for making deep learning models more interpretable and explainable. For example, techniques such as attention mechanisms [39] or saliency maps can highlight which parts of the input the model is focusing on to make its decision [40]. Hicks et al. [41] argued that deep learning predictions and decisions need to be accompanied by explanations so that the doctors responsible for the clinical decision-making process trust, understand, and validate such models [42-44]. To this end, they proposed a new method called electrocardiogram gradient class activation map, which produces explanations for the results of the electrocardiogram (EEG) analysis. Along the same line of EEG analysis, Khasawneh et al. [45,46] proposed treating the signal in a similar fashion to the clinicians and perform the signal inspection visually using deep learning object detection algorithms. In the context of language models, further explainability can be achieved via the following practices:

• Documentation: Developers can document how the model was trained, what data was used, what decisions were made in the training process, and what assumptions were made. Moreover, ethical standards can be developed to standardize the training process. This can help increase transparency and accountability.

• Auditing: Regular auditing of the model can help identify and correct biases, as well as provide insights into how the model is making decisions.

• Collaboration: Collaboration between developers, users, and experts in relevant fields can help increase transparency and ensure that the model is being used in an ethical and responsible manner.

While these approaches can help address the lack of transparency in deep learning, it is important to acknowledge that achieving full transparency may not be possible or desirable (e.g., proprietary or copyrighted/patented models) in all cases.

### 3.4 Over-reliance

Researchers, professionals, or students may become over-reliant on Chatbots, language models, and AI in general. Thus, they may neglect critical thinking, leading to errors and inaccuracies in their research, studies, or practice/work. Such over-reliance can happen in several ways, as in the following examples:

• Dataset selection: Researchers may rely too heavily on Chatbots to generate synthetic data or to augment existing datasets. This can be problematic if the generated data is biased or does not accurately reflect the real-world data.

• Hypothesis generation: Language models can generate hypotheses or research questions based on input from researchers. While this can be a useful tool for exploring new areas of research, researchers should be cautious not to rely too heavily on the model's suggestions without independent validation.

- Data analysis: Chatbots can be used to analyze and summarize large datasets. While this can save time and resources, researchers should be cautious not to rely too heavily on the model's output without independent verification.
- Model selection: Researchers may choose to use ChatGPT (or other AI language models) as their primary research tool, rather than as one tool among many. This can lead to over-reliance on the model's output and a failure to consider alternative hypotheses or methods.

Over-reliance can lead to several problems, including:

- Biases: As we discussed earlier, Chatbots and language models can be biased in various ways. If researchers rely too heavily on the model's output, they may unknowingly replicate or amplify these biases.
- Errors: ChatGPT (like all models) is not infallible. If researchers rely too heavily on the model's output, they may introduce errors or inaccuracies into their research.
- Over-generalization: ChatGPT is trained on a large corpus of text and may not accurately reflect the nuances and complexities of the real world. If researchers rely too heavily on the model's output, they may over-generalize or oversimplify their findings.

To avoid over-reliance on AI and language models, researchers should be cautious in their use of the model and should use it in conjunction with other research methods and tools. They should also be aware of the model's limitations and potential biases, and should take steps to mitigate these risks.

### 3.5 Ethical Concerns

ChatGPT can raise ethical concerns such as privacy violations and job displacement (i.e., involuntary job loss). ChatGPT can generate responses that may violate users' privacy, and the use of ChatGPTs in various industries may lead to job displacement. There are several ethical concerns associated with ChatGPT and other AI models. Here are a few examples:

- Bias: As we discussed earlier, language models can be biased in various ways, such as in the training data or algorithms used. These biases can lead to unfair or discriminatory outcomes, such as in employment, healthcare, or criminal justice.
- Privacy: Chatbots can generate highly personalized output based on input from users, which can raise privacy concerns. For example, if a user inputs sensitive information into the model (such as health or financial data), the model's output could reveal that information to others.
- Accountability: ChatGPT (and other AI models) can make decisions or generate output with significant consequences for individuals or society as a whole (e.g., in healthcare or criminal justice). If the model makes an incorrect or biased decision, it can be difficult to hold it accountable for its actions.
- Transparency: As we discussed earlier, deep learning can lack transparency, meaning that it is often difficult to understand how the model arrived at a particular output or decision. This lack of transparency can make it difficult to identify and correct biases or to hold the model accountable for its actions.

• Misuse: ChatGPT can be misused for nefarious purposes, such as generating fake news or propaganda. Moreover, academia is ringing the alarm bills about the great possibilities for cheating on academic assignments using language models and Chatbots, coupled with the lagging behind of cheating detections software on this problem. This can have serious consequences for individuals and society as a whole.

To address these ethical concerns, researchers, developers, and users of ChatGPT should prioritize ethical considerations throughout the model's development and use. This can include:

• Fairness: Ensuring that the model is trained on diverse and representative data, and that it does not unfairly discriminate against any particular group of people.

• Privacy: Ensuring that the model is used in a way that respects users' privacy and that sensitive data is protected.

• Accountability: Ensuring that there are mechanisms in place to hold the model accountable for its decisions and actions.

• Transparency: Ensuring that the model's output is transparent and explainable, so that users can understand how the model arrived at its decisions.

• Responsible use: Ensuring that the model is used in an ethical and responsible manner, and that it is not misused for nefarious purposes.

Each one of these considerations open several avenues for research. For example, the algorithms for calculating similarity scores and cheating detection need to be developed to take under consideration the availability of powerful Chatbots like ChatGPT.

## 4  Future Research Directions

### 4.1  Explainability

One of the most crucial research directions is to develop methods to make deep learning in general and language models in particular more explainable. Explainability refers to the ability to understand how a model arrived at its output or decision, and to be able to explain that process in a way that is understandable to humans. This will enable researchers to understand the logic behind the models' decisions and provide transparency in their output. Explainability is an important research direction for Chatbots, as it can help address concerns around transparency and accountability. Failing to address explainability has great ramifications on the adoption and regulatory certification of AI techniques [47,48]. For example, the European general data protection regulation explicitly requires decisions made in healthcare among other areas to be traceable and explainable [49]. Explainable AI (XAI) is a research avenue in AI that gaining a lot of attention driven by the real-life deployment requirements of AI-based systems. A recent survey by Bai et al. [50] presents the recent advancements toward achieving explainable AI in pattern recognition.

One approach to achieving explainability is through the use of attention mechanisms. Attention mechanisms allow the model to focus on certain parts of the input when generating its output. They can generate probability distributions relating to the input, which serve as indicators on the importance of features. By visualizing the attention weights for each part of the input [51], we can gain insight into which parts of the input the model is focusing on and how it is using that information to generate its output. However, attention mechanisms according to Liu et al. [52] may be unable to identify the polarity of the impact of individual features due to suppression effects.

Another approach is to use model-agnostic methods to explain the output of deep learning models. These methods aim to explain the model's output without needing to know the internal workings of the model itself. For example, one such method is LIME (Local Interpretable Model-agnostic Explanations) [53], which generates a simple, interpretable model that approximates the behavior of the original model on a local scale. Aditya and Pal [54] proposed further refinements to LIME using Shapley values used in game theory, which provide several advantages in terms of efficiency, consistency, and symmetry.

In addition to these approaches, there is ongoing research into developing new methods for explainability. For example, recent work has explored the use of counterfactual explanations, which aim to explain how the output of the model would have changed if certain parts of the input had been different [31]. Another area of research is developing methods to explain the output of black box models when the input is a sequence of events over time, such as in the case of medical records or financial transactions.

Overall, the goal of explainability research is to provide users with a better understanding of how language models in particular arrive at their output, and to provide mechanisms for identifying and correcting biases or errors in the model's decision-making. This is an important area of research for ensuring the ethical use of ChatGPTs in a variety of applications.

### 4.2 Bias Detection and Mitigation

Another research direction is to detect and mitigate bias in ChatGPTs. Researchers can develop methods to identify and eliminate bias from Chatbots by using techniques such as adversarial training [55]. This method fine tunes language models and deep learning models in general through the introduction of adversarial samples in the training set, which tends to increase the robustness and generalization of the model. Toward this end, several approaches and algorithms have been proposed in the literature, including adversarial training for large neural language models (ALUM) [56], generative adversarial training [57], attacking to training (A2T) [58], and large-margin classification [59]. Bias detection and mitigation are important steps in ensuring that language models are used ethically and fairly. Here are some approaches to bias detection and mitigation:

• Data collection: Bias can be introduced in the training data that is used to train ChatGPTs. One approach to reducing bias is to ensure that the training data is diverse and representative of the population that the model will be used on. This can involve careful selection of data sources and cleaning and preprocessing the data to remove any biases.

• Bias metrics: Once the model is trained, it is important to measure the extent of any bias that may be present. This can be done using various bias metrics, such as the disparate impact or statistical parity difference. These metrics can help identify areas of the model that may be more prone to bias.

• Mitigation strategies: Once bias has been identified, there are various strategies that can be used to mitigate it. One approach is to modify the training data to reduce bias, for example by oversampling underrepresented groups. Another approach is to modify the model itself, for example by adding constraints or penalties to the training process that encourage fairer predictions. Alternatively, post-processing techniques can be used to adjust the model's predictions to ensure fairness.

• Regular monitoring: Bias detection and mitigation is an ongoing process, and it is important to regularly monitor the model for any new sources of bias that may emerge. This can involve setting up automated monitoring systems that flag potential bias in real time, as well as regular audits of the model's performance.

### 4.3 Multimodal Integration

Researchers can explore the integration of ChatGPTs with other modalities such as images and videos to enhance their applications in fields such as education and healthcare. Multimodal integration is an important research direction for language models, as it involves combining multiple modalities of information, such as text, images, and audio, to generate more comprehensive and accurate outputs. Multimodal integration can help Chatbots better understand and respond to complex inputs, and can enable more natural and intuitive interactions between humans and machines.

One approach to multimodal integration in ChatGPTs is to use a multimodal transformer architecture [60], which incorporates multiple modalities of input into a single transformer model. This approach has been used in a number of applications, such as image captioning (e.g., XGPT [61]) and video question answering (e.g., AVQA [62]), with promising results. Another approach is to use multimodal fusion techniques to combine the outputs of separate models trained on different modalities [63]. For example, Zhu et al. used separate models for text and images, and then combined their outputs using a fusion approach based on self-attention [64]. In addition to these approaches, there is ongoing research into developing new methods for multimodal integration in language models. For example, recent work has explored the use of reinforcement learning to learn how to weight different modalities of input based on their relative importance [65]. Another area of research is developing methods to incorporate multimodal inputs that are not synchronous, such as when text and audio inputs are recorded separately [66]. In general, multimodal integration is an important research direction for language models, as it enables more flexible and powerful interactions with users and can improve the accuracy and comprehensiveness of the model's output.

### 4.4 Transfer Learning

Transfer learning is a research direction that involves using pre-trained models to perform specific tasks in various fields. Researchers can explore how ChatGPTs can be fine-tuned for specific applications in education, software engineering, healthcare, and marketing.

Transfer learning is a powerful technique that has been successfully applied to language models [66]. Transfer learning refers to the process of training a model on one task or domain, and then transferring that knowledge to a new task or domain [67]. In the case of Chatbots, transfer learning involves pre-training a model on a large corpus of text data, and then fine-tuning the model on a specific task or domain. Such method has several benefits for ChatGPTs. First, it can help address the issue of limited training data for specific tasks or domains, by allowing the model to leverage the knowledge it has gained from pre-training on large amounts of data.

Second, transfer learning can help reduce the computational cost of training a new model from scratch, as the pre-trained model can serve as a starting point for fine-tuning.

There are several approaches to transfer learning in ChatGPTs. One common approach is to use a pre-trained model such as GPT-2 or GPT-3, which have been trained on large amounts of diverse text data. The pre-trained model can then be fine-tuned on a specific task or domain, such as sentiment analysis or question answering, by further training the model on a smaller dataset specific to that task. Another approach is to use transfer learning to adapt a pre-trained model to a new language. This involves pre-training the model on a large corpus of text data in the new language, and then fine-tuning the model on specific tasks or domains in that language. Overall, transfer learning is a powerful technique for language models, as it enables the model to leverage knowledge gained from pre-training on large amounts of data, and can help address the issue of limited training data for specific tasks or domains.

## 5 Conclusions

ChatGPT and other advanced language models/chatbots are powerful disruptive tools that have the potential to revolutionize various fields such as education, software engineering, healthcare, and marketing. However, its drawbacks, such as plagiarism, bias and lack of transparency, need to be addressed, and researchers need to explore research directions such as explainability, bias detection and mitigation, multimodal integration, and transfer learning to ensure ChatGPTs are used responsibly and thoughtfully. The work in this paper surveyed the possible avenues where language models can positively or negatively contribute to that area, what possible changes need to be made to counter the negatives or misuse scenarios, and the future research directions necessary to achieve wide, effective, and proper deployment.

## References

1. Kung, T.H.; Cheatham, M.; Medenilla, A.; Sillos, C.; Leon, L.D.; Elepaño, C.; Madriaga, M.; Aggabao, R.; Diaz-Candido, G.; Maningo, J.; et al. Performance of ChatGPT on USMLE: Potential for AI-Assisted Medical Education Using Large Language Models **2022**. https://doi.org/10.1101/2022.12.19.22283643.

2. Weizenbaum, J. *Computer power and human reason*; W.H. Freeman: New York, NY, USA, 1976.

3. Wallace, R.S. The Anatomy of A.L.I.C.E. In *Parsing the Turing Test*; Springer Netherlands, 2007; pp. 181–210. https://doi.org/10.1007/978-1-4020-6710-5_13.

4. Carpenter, R. Cleverbot — cleverbot.com. https://www.cleverbot.com/, 2008. [Accessed 27-Feb-2023].

5. OpenAI. ChatGPT: Optimizing Language Models for Dialogue — openai.com. https://openai.com/blog/chatgpt/, 2022. [Accessed 27-Feb-2023].

6. Yang, Z.; Dai, Z.; Yang, Y.; Carbonell, J.; Salakhutdinov, R.R.; Le, Q.V. XLNet: Generalized Autoregressive Pretraining for Language Understanding. In Proceedings of the Advances in Neural Information Processing Systems; Wallach, H.; Larochelle, H.; Beygelzimer, A.; dÁlché Buc, F.; Fox, E.; Garnett, R., Eds. Curran Associates, Inc., 2019, Vol. 32.

7. Lepikhin, D.; Lee, H.; Xu, Y.; Chen, D.; Firat, O.; Huang, Y.; Krikun, M.; Shazeer, N.; Chen, Z. GShard: Scaling Giant Models with Conditional Computation and Automatic Sharding. *CoRR* **2020**, *abs/2006.16668*, http://xxx.lanl.gov/abs/2006.16668 [2006.16668].

8. Liu, Y.; Ott, M.; Goyal, N.; Du, J.; Joshi, M.; Chen, D.; Levy, O.; Lewis, M.; Zettlemoyer, L.; Stoyanov, V. RoBERTa: A Robustly Optimized BERT Pretraining Approach. *CoRR* **2019**, *abs/1907.11692*, http://xxx.lanl.gov/abs/1907.11692 [1907.11692].

9. Raffel, C.; Shazeer, N.; Roberts, A.; Lee, K.; Narang, S.; Matena, M.; Zhou, Y.; Li, W.; Liu, P.J. Exploring the Limits of Transfer Learning with a Unified Text-to-Text Transformer. *Journal of Machine Learning Research* **2020**, *21*, 1–67.

10. Tapalova, O.; Zhiyenbayeva, N. Artificial Intelligence in Education: AIEd for Personalised Learning Pathways. *Electronic Journal of e-Learning* **2022**, *20*, 639–653. https://doi.org/10.34190/ejel.20.5.2597.

11. Kumar, V.; Boulanger, D. Explainable Automated Essay Scoring: Deep Learning Really Has Pedagogical Value. *Frontiers in Education* **2020**, *5*. https://doi.org/10.3389/feduc.2020.572367.

12. Raychev, V.; Vechev, M.; Yahav, E. Code completion with statistical language models. In Proceedings of the Proceedings of the 35th ACM SIGPLAN Conference on Programming Language Design and Implementation. ACM, 2014. https://doi.org/10.1145/2594291.2594321.

13. Le, K.T.; Rashidi, G.; Andrzejak, A. A methodology for refined evaluation of neural code completion approaches. *Data Mining and Knowledge Discovery* **2022**, *37*, 167–204. https://doi.org/10.1007/s10618-022-00866-9.

14. Tu, Z.; Su, Z.; Devanbu, P. On the localness of software. In Proceedings of the Proceedings of the 22nd ACM SIGSOFT International Symposium on Foundations of Software Engineering. ACM, 2014. https://doi.org/10.1145/2635868.2635875.

15. Allamanis, M.; Brockschmidt, M.; Khademi, M. Learning to Represent Programs with Graphs, 2017. https://doi.org/10.48550/ARXIV.1711.00740.

16. Allamanis, M.; Barr, E.T.; Devanbu, P.; Sutton, C. A Survey of Machine Learning for Big Code and Naturalness. *ACM Computing Surveys* **2018**, *51*, 1–37. https://doi.org/10.1145/3212695.

17. Adlung, L.; Cohen, Y.; Mor, U.; Elinav, E. Machine learning in clinical decision making. *Med* **2021**, *2*, 642–665. https://doi.org/10.1016/j.medj.2021.04.006.

18. Wu, L.; Chen, S.; Guo, L.; Shpyleva, S.; Harris, K.; Fahmi, T.; Flanigan, T.; Tong, W.; Xu, J.; Ren, Z. Development of benchmark datasets for text mining and sentiment analysis to accelerate regulatory literature review. *Regulatory Toxicology and Pharmacology* **2023**, *137*, 105287. https://doi.org/10.1016/j.yrtph.2022.105287.

19. Lederman, A.; Lederman, R.; Verspoor, K. Tasks as needs: reframing the paradigm of clinical natural language processing research for real-world decision support. *Journal of the American Medical Informatics Association* **2022**, *29*, 1810–1817. https://doi.org/10.1093/jamia/ocac121.

20. Gururangan, S.; Marasovic, A.; Swayamdipta, S.; Lo, K.; Beltagy, I.; Downey, D.; Smith, N.A. Don't Stop Pretraining: Adapt Language Models to Domains and Tasks. *CoRR* **2020**, *abs/2004.10964*, http://xxx.lanl.gov/abs/2004.10964 [2004.10964].

21. Liu, Z.; Roberts, R.A.; Lal-Nag, M.; Chen, X.; Huang, R.; Tong, W. AI-based language models powering drug discovery and development. *Drug Discovery Today* **2021**, *26*, 2593–2607. https://doi.org/10.1016/j.drudis.2021.06.009.

22. Tripathi, A.; Misra, K.; Dhanuka, R.; Singh, J.P. Artificial Intelligence in Accelerating Drug Discovery and Development. *Recent Patents on Biotechnology* **2023**, *17*, 9–23. https://doi.org/10.2174/1872208316666220802151129.

23. Bhatnagar, R.; Sardar, S.; Beheshti, M.; Podichetty, J.T. How can natural language processing help model informed drug development?: a review. *JAMIA Open* **2022**, *5*. https://doi.org/10.1093/jamiaopen/ooac043.

24. Leyens, L.; Reumann, M.; Malats, N.; Brand, A. Use of big data for drug development and for public and personal health and care. *Genetic Epidemiology* **2016**, *41*, 51–60. https://doi.org/10.1002/gepi.22012.

25. Ball, R.; Pan, G.D. "Artificial Intelligence" for Pharmacovigilance: Ready for Prime Time? *Drug Safety* **2022**, *45*, 429–438. https://doi.org/10.1007/s40264-022-01157-4.

26. Koneti, G.; Das, S.S.; Bahl, J.; Ranjan, P.; Ramamurthi, N. Discovering the Knowledge in Unstructured Early Drug Development Data Using NLP and Advanced Analytics. In Proceedings of the 2022 IEEE International Conference on Bioinformatics and Biomedicine (BIBM), 2022, pp. 3840–3842.

https://doi.org/10.1109/BIBM55620.2022.9995435.

27. Wang, X.; Xu, X.; Tong, W.; Liu, Q.; Liu, Z. DeepCausality: A general AI-powered causal inference framework for free text: A case study of LiverTox. *Frontiers in Artificial Intelligence* **2022**, *5*. https://doi.org/10.3389/frai.2022.999289.

28. Lee, J.; Yoon, W.; Kim, S.; Kim, D.; Kim, S.; So, C.H.; Kang, J. BioBERT: a pre-trained biomedical language representation model for biomedical text mining. *Bioinformatics* **2019**, *36*, 1234–1240. https://doi.org/10.1093/bioinformatics/btz682.

29. Huang, K.; Altosaar, J.; Ranganath, R. ClinicalBERT: Modeling Clinical Notes and Predicting Hospital Readmission. *CoRR* **2019**, *abs/1904.05342*, http://xxx.lanl.gov/abs/1904.05342 [1904.05342].

30. Alsentzer, E.; Murphy, J.; Boag, W.; Weng, W.H.; Jindi, D.; Naumann, T.; McDermott, M. Publicly Available Clinical. In Proceedings of the Proceedings of the 2nd Clinical Natural Language Processing Workshop. Association for Computational Linguistics, 2019. https://doi.org/10.18653/v1/w19-1909.

31. Verma, S.; Sharma, R.; Deb, S.; Maitra, D. Artificial intelligence in marketing: Systematic review and future research direction. *International Journal of Information Management Data Insights* **2021**, *1*, 100002. https://doi.org/10.1016/j.jjimei.2020.100002.

32. Nguyen, Q.N.; Sidorova, A.; Torres, R. User interactions with chatbot interfaces vs. Menu-based interfaces: An empirical study. *Computers in Human Behavior* **2022**, *128*, 107093. https://doi.org/10.1016/j.chb.2021.107093.

33. Chipman, S. What is CRM artificial intelligence and what can it do for my business? - CRM Switch — crmswitch.com, 2023. [Accessed 06-Mar-2023].

34. Mauro, A.D.; Sestino, A.; Bacconi, A. Machine learning and artificial intelligence use in marketing: a general taxonomy. *Italian Journal of Marketing* **2022**, *2022*, 439–457. https://doi.org/10.1007/s43039-022-00057-w.

35. Duarte, V.; Zuniga-Jara, S.; Contreras, S. Machine Learning and Marketing: A Systematic Literature Review. *IEEE Access* **2022**, *10*, 93273–93288. https://doi.org/10.1109/access.2022.3202896.

36. Bruyn, A.D.; Viswanathan, V.; Beh, Y.S.; Brock, J.K.U.; von Wangenheim, F. Artificial Intelligence and Marketing: Pitfalls and Opportunities. *Journal of Interactive Marketing* **2020**, *51*, 91–105. https://doi.org/10.1016/j.intmar.2020.04.007.

37. Schramowski, P.; Turan, C.; Andersen, N.; Rothkopf, C.A.; Kersting, K. Large pre-trained

language models contain human-like biases of what is right and wrong to do. *Nature Machine Intelligence* **2022**, *4*, 258–268. https://doi.org/10.1038/s42256-022-00458-8.

38. Rudin, C. Stop explaining black box machine learning models for high stakes decisions and use interpretable models instead. *Nature Machine Intelligence* **2019**, *1*, 206–215. https://doi.org/10.1038/s42256-019-0048-x.

39. Yang, X. An Overview of the Attention Mechanisms in Computer Vision. *Journal of Physics: Conference Series* **2020**, *1693*, 012173. https://doi.org/10.1088/1742-6596/1693/1/012173.

40. An, J.; Joe, I. Attention Map-Guided Visual Explanations for Deep Neural Networks. *Applied Sciences* **2022**, *12*, 3846. https://doi.org/10.3390/app12083846.

41. Hicks, S.A.; Isaksen, J.L.; Thambawita, V.; Ghouse, J.; Ahlberg, G.; Linneberg, A.; Grarup, N.; Strümke, I.; Ellervik, C.; Olesen, M.S.; et al. Explaining deep neural networks for knowledge discovery in electrocardiogram analysis. *Scientific Reports* **2021**, *11*. https://doi.org/10.1038/s41598-021-90285-5.

42. Riegler, M.; Lux, M.; Griwodz, C.; Spampinato, C.; de Lange, T.; Eskeland, S.L.; Pogorelov, K.; Tavanapong, W.; Schmidt, P.T.; Gurrin, C.; et al. Multimedia and Medicine. In Proceedings of the Proceedings of the 24th ACM international conference on Multimedia. ACM, 2016. https://doi.org/10.1145/2964284.2976760.

43. Kelly, C.J.; Karthikesalingam, A.; Suleyman, M.; Corrado, G.; King, D. Key challenges for delivering clinical impact with artificial intelligence. *BMC Medicine* **2019**, *17*. https://doi.org/10.1186/s12916-019-1426-2.

44. Chen, D.; Liu, S.; Kingsbury, P.; Sohn, S.; Storlie, C.B.; Habermann, E.B.; Naessens, J.M.; Larson, D.W.; Liu, H. Deep learning and alternative learning strategies for retrospective real-world clinical data. *npj Digital Medicine* **2019**, *2*. https://doi.org/10.1038/s41746-019-0122-0.

45. Khasawneh, N.; Fraiwan, M.; Fraiwan, L. Detection of K-complexes in EEG waveform images using faster R-CNN and deep transfer learning. *BMC Medical Informatics and Decision Making* **2022**, *22*. https://doi.org/10.1186/s12911-022-02042-x.

46. Khasawneh, N.; Fraiwan, M.; Fraiwan, L. Detection of K-complexes in EEG signals using deep transfer learning and YOLOv3. *Cluster Computing* **2022**. https://doi.org/10.1007/s10586-022-03802-0.

47. Samek, W.; Montavon, G.; Vedaldi, A.; Hansen, L.K.; Müller, K.R.  *Explainable AI: Interpreting, explaining and visualizing deep learning*; Springer Nature: Cham, Switzerland, 2019.

48. Samek, W.; Müller, K.R.  Towards Explainable Artificial Intelligence. In *Explainable AI: Interpreting, Explaining and Visualizing Deep Learning*; Springer International Publishing, 2019; pp. 5–22.  https://doi.org/10.1007/978-3-030-28954-6_1.

49. Singh, A.; Sengupta, S.; Lakshminarayanan, V.  Explainable Deep Learning Models in Medical Image Analysis.  *Journal of Imaging* **2020**, *6*, 52.  https://doi.org/10.3390/jimaging6060052.

50. Bai, X.; Wang, X.; Liu, X.; Liu, Q.; Song, J.; Sebe, N.; Kim, B.  Explainable deep learning for efficient and robust pattern recognition: A survey of recent developments.  *Pattern Recognition* **2021**, *120*, 108102.  https://doi.org/10.1016/j.patcog.2021.108102.

51. Choo, J.; Liu, S.  Visual Analytics for Explainable Deep Learning.  *IEEE Computer Graphics and Applications* **2018**, *38*, 84–92.  https://doi.org/10.1109/MCG.2018.042731661.

52. Liu, Y.; Li, H.; Guo, Y.; Kong, C.; Li, J.; Wang, S.  Rethinking Attention-Model Explainability through Faithfulness Violation Test.  *CoRR* **2022**, *abs/2201.12114*, http://xxx.lanl.gov/abs/2201.12114 [2201.12114].

53. Ribeiro, M.T.; Singh, S.; Guestrin, C.  "Why Should I Trust You?": Explaining the Predictions of Any Classifier.  *CoRR* **2016**, *abs/1602.04938*, http://xxx.lanl.gov/abs/1602.04938 [1602.04938].

54. Aditya, P.S.R.; Pal, M.  Local Interpretable Model Agnostic Shap Explanations for machine learning models, 2022.  https://doi.org/10.48550/ARXIV.2210.04533.

55. Zhao, W.; Alwidian, S.; Mahmoud, Q.H.  Adversarial Training Methods for Deep Learning: A Systematic Review.  *Algorithms* **2022**, *15*, 283.  https://doi.org/10.3390/a15080283.

56. Liu, X.; Cheng, H.; He, P.; Chen, W.; Wang, Y.; Poon, H.; Gao, J.  Adversarial Training for Large Neural Language Models.  *CoRR* **2020**, *abs/2004.08994*, http://xxx.lanl.gov/abs/2004.08994 [2004.08994].

57. Movahedi, S.; Shakery, A. Generative Adversarial Training Can Improve Neural Language Models, 2022. https://doi.org/10.48550/ARXIV.2211.09728.

58. Yoo, J.Y.; Qi, Y. Towards Improving Adversarial Training of NLP Models. In Proceedings of the Findings of the Association for Computational Linguistics: EMNLP 2021; Association for Computational Linguistics: Punta Cana, Dominican Republic, 2021; pp. 945–956. https://doi.org/10.18653/v1/2021.findings-emnlp.81.

59. Wang, D.; Gong, C.; Liu, Q. Improving Neural Language Modeling via Adversarial Training. In Proceedings of the Proceedings of the 36th International Conference on Machine Learning; Chaudhuri, K.; Salakhutdinov, R., Eds. PMLR, 2019, Vol. 97, *Proceedings of Machine Learning Research*, pp. 6555–6565.

60. Xu, P.; Zhu, X.; Clifton, D.A. Multimodal Learning with Transformers: A Survey, 2022. https://doi.org/10.48550/ARXIV.2206.06488.

61. Xia, Q.; Huang, H.; Duan, N.; Zhang, D.; Ji, L.; Sui, Z.; Cui, E.; Bharti, T.; Zhou, M. XGPT: Cross-modal Generative Pre-Training for Image Captioning. In *Natural Language Processing and Chinese Computing*; Springer International Publishing, 2021; pp. 786–797. https://doi.org/10.1007/978-3-030-88480-2_63.

62. Yun, H.; Yu, Y.; Yang, W.; Lee, K.; Kim, G. Pano-AVQA: Grounded Audio-Visual Question Answering on 360 Videos. In Proceedings of the 2021 IEEE/CVF International Conference on Computer Vision (ICCV). IEEE, 2021. https://doi.org/10.1109/iccv48922.2021.00204.

63. Pawłowski, M.; Wróblewska, A.; Sysko-Romańczuk, S. Effective Techniques for Multimodal Data Fusion: A Comparative Analysis. *Sensors* **2023**, *23*, 2381. https://doi.org/10.3390/s23052381.

64. ] Zhu, H.; Wang, Z.; Shi, Y.; Hua, Y.; Xu, G.; Deng, L. Multimodal Fusion Method Based on Self-Attention Mechanism. *Wireless Communications and Mobile Computing* **2020**, *2020*, 1–8. https://doi.org/10.1155/2020/8843186.

65. Bernardino, G.; Jonsson, A.; Loncaric, F.; Castellote, P.M.M.; Sitges, M.; Clarysse, P.; Duchateau, N. Reinforcement Learning for Active Modality Selection During Diagnosis. In *Lecture Notes in Computer Science*; Springer Nature Switzerland, 2022; pp. 592–601.

https://doi.org/10.1007/978-3-031-16431-6_56.

66. Bayoudh, K.; Knani, R.; Hamdaoui, F.; Mtibaa, A. A survey on deep multimodal learning for computer vision: advances, trends, applications, and datasets. *The Visual Computer* **2021**, *38*, 2939–2970. https://doi.org/10.1007/s00371-021-02166-7.

67. Fraiwan, M.; Audat, Z.; Fraiwan, L.; Manasreh, T. Using deep transfer learning to detect scoliosis and spondylolisthesis from x-ray images. *PLOS ONE* **2022**, *17*, e0267851. https://doi.org/10.1371/journal.pone.0267851.